\documentclass[prl,aps,amssymb,twocolumn,showpacs,superscriptaddress]{revtex4}
\usepackage{graphicx}

\begin{document}

\title{Full Counting Statistics of Avalanche Transport: an Experiment}

\author{J.~Gabelli}
\author{B.~Reulet}
\affiliation{Laboratoire de Physique des Solides, UMR8502, b\^atiment 510, Universit\'e Paris-Sud, 91405 ORSAY Cedex, France}

\date{\today}
\begin{abstract}
We report the first measurement of high order cumulants of the current fluctuations in an avalanche diode run through by a stationary dc current. Such a system is archetypic of devices in which transport is governed by a collective mechanism, here charge multiplication by avalanche. We have measured the first 5 cumulants of the probability distribution of the current fluctuations. We show that the charge multiplication factor is distributed according to a power law that is different from that of the usual avalanche below breakdown, when avalanches are well separated.
\end{abstract}
 \maketitle

Current noise, i.e. the variance of the current fluctuations, is the simplest measure of the statistical aspect of electronic transport in a conductor, beyond the dc current. Its study as a function of other parameters (voltage, temperature, etc.) has been a powerful way to check our understanding of the conduction process in many systems, and a tool to obtain information that is hidden in the mean current \cite{BuBlan}.

In order to probe the statistics of the conduction more in depth, a better knowledge of the distribution function $P$ of the fluctuating current $I(t)$ is necessary, beyond the average $\langle I\rangle$, the dc current,  and the variance $\langle i^2\rangle$ with $i(t)=I(t)-\langle I\rangle$. The brackets $\langle.\rangle$ denote average over the distribution $P$. Most of the time, the full measurement of $P(i)$ is not possible, but a finite number of the moments $\langle i^n \rangle$ of the distribution can be measured, which give some insight into the statistics of $i$. For example, the third moment $\langle i^3 \rangle$ reveals the asymmetry (skewness)of the distribution around the average.

Triggered by the calculation of the full counting statistics of current fluctuations in a quantum conductor  \cite{LevFCS}, the measurement of higher moments of current fluctuations beyond the second has started only recently \cite{S3BR}. This measurement and later work \cite{Bomze} have confirmed the Poissonian aspect of transport in a tunnel junction, by proving that the spectral density of the third moment of the current fluctuations is $S_3=e^2\langle I\rangle$ (where $e$ is the electron charge), independent of temperature. After that, measurements of higher moments have been achieved in Coulomb blockaded quantum dots \cite{Leturcq, Flimt}. In all these measurements, the statistics of transport is driven by the finite rate at which electrons can pass a barrier.

In this article we report the first measurement of high order cumulants of current fluctuations in avalanche diodes. In such samples, the process that is responsible for current fluctuations is not the transport at the one electron level, but arises from a collective phenomenon, charge multiplication. This occurs due to spontaneous creation of electron-hole pairs in semi-conductors in the presence of a high electric field, which strongly accelerates the charge carriers (electrons or holes). When their kinetic energy reaches the gap of the semi-conductor, they may give part of it to create an new electron-hole (e-h) pair which, in turn, will be accelerated and give birth to other e-h pairs. Thus, each charge entering the sample generates a current pulse of total charge $Me$ with $M\gg1$. This avalanche is a statistical process, since the e-h pair creation occurs only with a finite probability per unit time: $M$ is not a well defined number but has a very wide statistical distribution. This distribution has been studied both theoretically and experimentally in the regime where no current is injected but individual e-h pairs are photo-created, giving rise to well separated current pulses \cite{Tager, McIntyre, Conradi72}. In such a regime, adequate for the use of avalanche process as an amplifier of individual events, like in the avalanche photodiode, the statistics of the pulses determines the noise added by the detector. We are not considering here this regime which is not stationary, but our goal is to investigate the current noise due to avalanche process in the so called breakdown regime, when the sample sustains a stationary dc current. Semi-conductors working in this regime are indeed used as noise sources.

\begin{figure}
\includegraphics[width= 0.85\columnwidth]{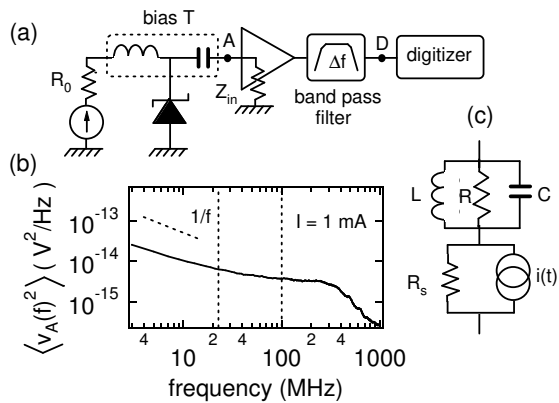}
\caption{(a) Schematics of the experimental setup. $Z_{in}$ is the input impedance of the amplifier, that can be modified. (b) Spectral density of voltage fluctuations at point $A$ vs. frequency for $I=1\,$mA. (c) Equivalent circuit of the sample. $R_s=4.6\,\Omega$, $L=12.6\,$nH, $C=5.1\,$pF, $R[I=1mA]=500\,\Omega$. $i(t)$ is the current source that models the avalanche.}
\label{figschem}
\end{figure}

\emph{Experimental setup --- } The sample we have studied is a SM-1 avalanche diode manufactured by Micronetics\cite{Micronetics} (we have measured several devices, all showing very similar features). The avalanche starts when the component is reverse biased by a voltage $\gtrsim 8$V. Too close to this value there is no stationary regime. We will focus only on the bias region that corresponds to a dc current $I\geq0.46$ mA  where we observe a stationary regime (in the following $I=\langle I\rangle$ stands for the dc current). The schematics of the biasing and measurement circuit is depicted in Fig. \ref{figschem}(a). We use a bias tee to separate the dc part of the current (up to 100 kHz) from its fluctuations (100 kHz to 1 GHz). The sample is biased by applying a dc voltage through a $R_0=1\;\mathrm{k}\Omega$ resistor in series with the inductor of the bias tee. The fluctuating voltage across the sample is detected through the capacitive part of the bias tee by a spectrum analyzer (at point A) or amplified (in the range 1- 500 MHz), filtered and converted into a digital signal (at point D) by a 14 bits, 200MS/s A to D converter. All the measurements have been performed at room temperature.

A spectrum analyzer measures the power spectral density of the voltage as a function of frequency. This quantity is proportional to the noise generated by the sample at frequency $f$ within a bandwidth of 1Hz. A  spectrum for a bias of $I=1\,$mA is shown in Fig. \ref{figschem}(b). Three regions can be distinguished. At low frequency, the noise spectral density decays with increasing frequency, approximately like $1/f$. At higher frequency, the spectrum is almost white, up to $\sim500\,$MHz. In order to remove the contribution of the $1/f$ noise (unrelated to the physics of avalanche \cite{1overf}) and in order to avoid aliasing, we have used a $25-100\,$MHz bandpass filter before digitizing the signal (indicated by vertical dashed lines in Fig. \ref{figschem}(b)).

In our setup, the voltage $v_D(t)$ recorded by the digitization card is related to the intrinsic current source $i(t)$ that models the avalanche process by $v_D(f)=H(f)i(f)$ where $v_D(f)$ and $i(f)$ are the Fourier components at frequency $f$ of $v_D(t)$ and $i(t)$. The transfer function $H(f)$ contains the effects of the impedances and gains of the external circuit (bias tee, amplifier, filter) as well as the parasitic impedances inside the component. We have measured the frequency response of the setup as well as the complex impedance $Z_s$ of the sample as a function of frequency and dc voltage with a vector network analyzer in the range $1-1000\,$MHz. We observe that the sample can be well modeled by a $R_s=4.6\,\Omega$ resistor in series with a parallel RLC resonator, see Fig. \ref{figschem}(c). $R_s$, $L$ and $C$ are independent of the bias voltage, but $R$ varies between $250\,\Omega$ and $750\,\Omega$ when $I$ varies between $0.46\,$mA and $4.39\,$mA, which implies that $H(f)$ depends on the bias voltage too. One also has to know where to incorporate the current source $i(t)$ created by the avalanche in the schematics of the sample. Knowing that the avalanche itself has a white spectrum \cite{Tager}, we conclude from the frequency dependence of the noise spectral density (Fig. \ref{figschem}(b)) that $i(t)$ must be in parallel with $R_s$, see Fig. \ref{figschem}(c). The high frequency roll-off observed on the power spectrum comes from the RLC circuit ($1/(2\pi\sqrt{LC})=600\,$MHz). Thus, a direct determination of $i(t)$ could be obtained by dividing by $R_s$ the voltage across the diode, measured  with a high impedance voltmeter. We have used an amplifier with $50\,\Omega$ input impedance and calculated $H(f)$ (which includes the finite input impedance of the amplifier) to deduce $i(f)$ from $v_D(f)$.

\begin{figure}
\includegraphics[width= 0.85\columnwidth]{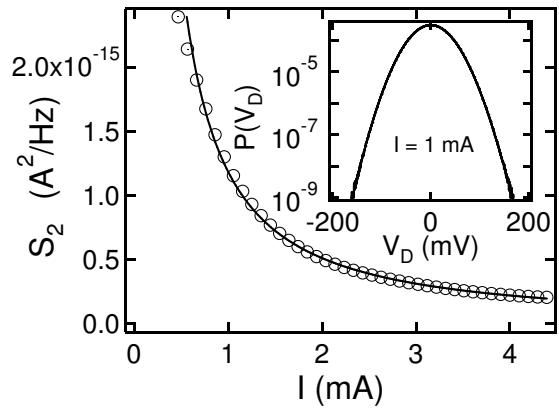}
\caption{Noise spectral density $S_2(f)=\langle|i(f)|^2\rangle$, vs. dc current (symbols: measurement, solid line: fit $I^{-1.21}$). Inset: measured probability distribution of $v_D$ for $I=1\,$mA.}
\label{figS2}
\end{figure}

\emph{Second moment -- } We have recorded the voltage fluctuations $v_D(t)$ and made histograms of them for many bias points, see inset of Fig. \ref{figS2}. The variance $\langle v_D^2\rangle$ (here the brackets mean time averaging) is related to the noise spectral density $S_2(f)=\langle|i(f)|^2\rangle$ by: $\langle v_D^2\rangle=\int_{-\infty}^{+\infty} |H(f)|^2 S_2(f)df $.
Since we observe that $S_2(f)$ is reasonably frequency independent within the detection bandwidth (see Fig. \ref{figschem}(b)), the variance of $v_D$ and the spectral density of the current fluctuations $S_2$ are simply proportional: $\langle v_D^2\rangle=H_2S_2$ where $H_2(V)=\int |H(V,f)|^2df$. We show on Fig. \ref{figS2} the variations of $S_2$ with the dc current running through the sample. Note that the amount of noise emitted by the avalanche diode is huge, so that the contribution of the amplifier is negligible. For $I=1\,$mA, $S_2=1.15\times10^{-15}\, \mathrm{A^2/Hz}$ which is equivalent to the Johnson noise of a $50\,\Omega$ resistor at a temperature of 18 millions Kelvin degrees, to compare with the $300\,$K noise temperature of the amplifier. Defining the Fano factor $F_2$ by $S_2=F_2eI$, we find $F_2=6.8\times10^6$ for $I=1\,$mA. The noise is thus extremely super-Poissonian ($F_2=1$ for Poisson noise).

\begin{figure}
\includegraphics[width= 0.9\columnwidth]{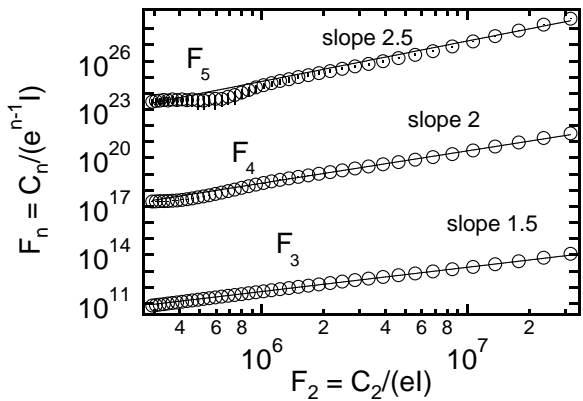}
\caption{Fano factors $F_n=C_n/(e^{n-1}I)$ of the first cumulants of the distribution of current fluctuations, as the function of the second Fano factor $F_2=C_2/(eI)$, in log-log scale.} \label{figCn}
\end{figure}

\emph{Third moment and environmental effects --- } From the histograms of $v_D(t)$ we have calculated the third moment $\langle v_D^3\rangle$. It is related to spectral density $S_3(f_1,f_2)=\langle i(f_1)i(f_2)i(-f_1-f_2)\rangle$ of the third moment of $i(t)$ by \cite{BRSPIE}: $\langle v_D^3\rangle=\!\int\!\!\!\int H(f_1)H(f_2)H(-f_1-f_2) S_3(f_1,f_2) df_1df_2$.
If we suppose that like $S_2$, $S_3$ is frequency independent \cite{noteS3freq}, one has: $\langle v_D^3\rangle=H_3S_3$ with $H_3=\int\int df_1df_2H(f_1)H(f_2)H(-f_1-f_2)$. As it has been demonstrated experimentally and well understood \cite{S3BR,BRSPIE,Beenakker}, there is an additional contribution to $S_3$ given by: $R_{eff}S_2\frac{dS_2}{dV}$. $R_{eff}$ is an effective resistor that contains the impedance of the sample and that of the environment, here the internal RLC circuit and the input impedance of the amplifier $Z_{in}$. In order to estimate the environmental contribution, we have modified $R_{eff}$ by adding resistances to ground at the input of the amplifier, thus varying $Z_{in}$ in the range $4-50\,\Omega$. This had no effect on $S_3$ (besides an overall reduction due to voltage division by the added resistances) except at very low current, where both $S_2$ and $|dS_2/dV|$ are maximum, and where we observe a 5\% negative deviation, in agreement with the prediction. Thus, environmental contributions to $S_3$ are negligible, i.e. we have access to the intrinsic third moment of the current fluctuations. Moreover, $S_3$ (and higher moments) not being influenced by $Z_{in}$ demonstrates that the statistics of the current is not affected by the non-linearity of the component (which is small since $L\omega\ll R$ in our detection range). We have plotted the result for $S_3(I)$ in Fig. \ref{figCn}, as $F_3=S_3/(e^2I)$ vs. $F_2=S_2/(eI)$. We discuss the interpretation of this result later.

\begin{figure}
\includegraphics[width= 0.85\columnwidth]{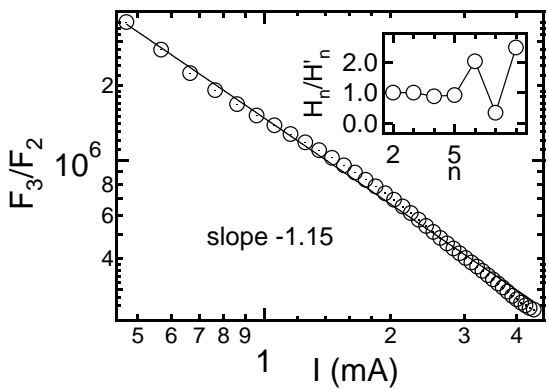}
\caption{Ratio of Fano factors $F_3/F_2=C_3/(eC_2)$ as a function of the dc current $I$, in log-log scale. Inset: ratio of two independent sets of calibration coefficients for the cumulants.}
\label{figm2}
\end{figure}

\emph{Higher order cumulants --- }
Any statistical measurement has to face a fundamental problem: the central limit theorem, according to which the distribution of the sum of $N$ independent variables tends to a Gaussian when $N$ goes to infinity, regardless of the distribution law of each of the variables. In our measurement, a current $I=1\,$mA integrated during $5\,$ns corresponds to $N\approx3\times10^7$ electrons, and the measured histograms are very close to gaussians, see inset of Fig. \ref{figS2}. Nevertheless one can access the probability distribution $P(i)$ by working with its cumulants, defined as follows. We introduce the characteristic function  $\chi(z)=\langle e^{iz}\rangle$ ($i$ is the fluctuating current). The Taylor expansion of $\chi(z) =\sum_n \langle i^n\rangle z^n/n!$ provides the moments $\langle i^n\rangle$. The cumulants $\langle\langle i^n\rangle\rangle$ are defined by the Taylor expansion of $\ln\chi(z)=\sum_n \langle\langle i^n\rangle\rangle z^n /n!$. If $P(i)$ is Gaussian, $\ln\chi(z)$ is a second degree polynomial, and $\langle\langle i^n\rangle\rangle=0$ for $n\geq3$. Thus the cumulants reveal how much a distribution deviates from a Gaussian. From the histograms of $v_D(t)$ we calculate the cumulants $\langle\langle v_D^n\rangle\rangle$ from which we deduce the spectral cumulants of $P(i)$, which we note $C_n$, supposing that they are frequency independent, as $C_2=S_2$ and $C_3=S_3$ are. A simple calculation gives indeed: $\langle\langle v_D^n\rangle\rangle=H_nC_n$ where the coefficients $H_n$ are given by: $ H_n=\int_0^{+\infty} H^n(t) dt $
with $H(t)$ the inverse Fourier transform of $H(f)$, i.e. the impulse response of the circuit. This definition coincides with that given before for $n=2$ and $n=3$. In order to check the calibration coefficients $H_n$, we have measured the 5 first cumulants of a pseudo-random current generated by a 1GS/s arbitrary waveform generator with a statistics that we have computed. The ratio of the measured cumulants to the real ones provides us with another set of calibration coefficients $H_n'$. We obtain $H_n\simeq H_n'$ up to $n=5$, see inset of Fig. \ref{figm2}. The systematic error on the sixth spectral cumulant is only by a factor 2. We conclude that we can perform a reliable and quantitative measurement of $C_n$ up to $n=5$.

We have measured the 5 first spectral cumulants of the current noise of the avalanche diode. We observe that they all decrease as a function of the average current as a non-integer power law (see e.g. $S_2$ on Fig. \ref{figS2}). In order to synthesize our results, we show on Fig. \ref{figCn} the Fano factors $F_n$ of the n$^\mathrm{th}$ cumulant ($n\geq3$) as a function of the second one $F_2$; we define the (dimensionless) Fano factors by: $C_n=F_n e^{n-1}I$. We observe a remarkable behavior: $F_n(I)\propto F_2^{n/2}(I)$, which can be rewritten as $F_n(I)=g_n G^n(I)$  ($n\geq2$) where the function $G(I)$ is the same for all the cumulants and $g_n$ are constants.
Note that all the cumulants can be in principle calculated from the data, but the uncertainty on $C_n$ due to the statistical aspect of the measurement (we record $v_D(t)$ during a finite time) grows fast with $n$ \cite{JYO}. Fig. \ref{figCn} corresponds to $8\times10^{10}$ independent measurements of the current for each bias point.

\emph{Interpretation --- }
The result of Fig. \ref{figCn} is central in the understanding of the statistical aspect of transport in the avalanche regime. A Poisson distribution with an effective charge $q=Me$, has Fano factors $F_n=M^{n-1}=F_2^{n-1}$, even if the effective charge depends on $I$. Our measurements clearly rules out this scenario. The reason for this is that the avalanche itself is responsible for the current noise.
We will now compare our experimental result with what is predicted in the non-stationnary regime. An injected charge  gives rise to $m$ charges with a probability $P(m)$. Thus, an injected photo-current $i_0$ gives rise to a dc current $I=\langle m\rangle i_0$ and to current fluctuations of spectral cumulants $C_n=e^{n-1}i_0\langle\langle m^p\rangle\rangle$ \cite{McIntyre,VanVliet}. The multiplication factor $m$ has been shown to obey a probability distribution such that: $\langle\langle m^n\rangle\rangle\sim\langle m^n\rangle\propto \langle m\rangle^{2n-1}$ \cite{McIntyre,VanVliet}. This implies for the Fano factors: $F_n\propto \langle m\rangle^{2n-2}\propto F_2^{n-1}$, in strong contrast with our results.

The non-stationnary regime is a particular case of avalanches, which are more generally described by a probability distribution that is a power law, $P(m)\sim m^{-\tau}$ (with $1<\tau<2$) valid for $m_1\sim1\ll m \ll m_2$ \cite{LeDoussal}. The results of the non-stationnary regime are recovered taking $\tau=3/2$ \cite{McIntyre}, the mean field result of theory of avalanches . $m_1$ represents the most probable avalanche, which is small, and $m_2$ the largest one, which is somewhat rare but dominates the statistics of $m$. Indeed, $\langle\langle m^p\rangle\rangle\sim\langle m^p\rangle\propto m_2^{p+1-\tau}$. The average multiplication factor is $\langle m \rangle\simeq m_2^{2-\tau}\ll m_2$, and the current spectral cumulants $C_p=k_p e^{p-1}i_0 m_2^{p+1-\tau}$. The coefficients $k_p$ depend on the exact shape of $P(m)$ for $m\leq m_1$ and $m\geq m_2$. This describes our experimental results provided $I\propto i_0 m_2^{1-\tau}$. Then the function $G(I)$ is simply $G(I)=m_2$. However, in the stationary regime, in which our experiments have been performed, the dc current is imposed by a battery, and not related to any $i_0$, which should disappear from the theory. This implies a power law relation between the average current and the largest avalanche $m_2$: $I\propto m_2^{1-\tau}$, which reflects how the probability distribution of the multiplication factor adapts to the applied dc current. From our experimental data we can extract the current dependence of $m_2$ through $F_3/F_2=(k_3/k_2) m_2$. As can be seen in Fig. \ref{figm2}, we indeed observe a power law, from which we deduce $\tau\simeq1.87$. The largest avalanche is huge, $m_2=1.38\times10^6$ for $I=1\,$mA and $k_2=k_3$, whereas the average avalanche is modest, $\langle m\rangle=6.3$. From $F_n=k_n m_2^n$ and taking $k_2=k_3$ we deduce $k_2=3.6\times10^{-6}$, $k_4=8.6k_2$ and $k_5=20.9k_2$. A detailed theory is of course needed, in particular to explain how the avalanche feeds itself to reach the stationary regime.

In conclusion, we have demonstrated how the statistical physics of a complex phenomenon can be extracted from the analysis of its fluctuations beyond the second moment. In particular, we have demonstrated that the charge multiplication factor of avalanche diodes has a power law probability distribution different from what has been reported until now in the non-stationary regime, and different from what is predicted using mean field approximation in avalanche theories. This may open new routes for the study of other components, or complex conductors such as those with charge density waves. Moreover, avalanche diodes are advertised as perfect gaussian noise sources. This is definitely incorrect and may have practical consequences. For example, using the sign of the instantaneous fluctuating current as a source of random bits gives, for the sample we have analyzed and in our bandwidth, $P(i>0) \simeq 1.01 P(i<0)$. To cure this, the central limit theorem may help, with the price of a smaller bit rate.

We are very grateful to P. Matthews from Micronetics for providing us with noise diodes. We thank M. Aprili, W. Belzig, P. Le Doussal, T. Novotn\'{y}, J.-Y. Ollitrault and K. Wiese for fruitful discussions. This work was supported by ANR-05-NANO-039-02 and by Triangle de la Physique.



\begin{thebibliography}{99}
\bibitem{noteS3freq} We have calculated the correlation function $\langle v_D(t) v_D(t+\tau_1) v_D(t+\tau_2) \rangle$, proportional to the Fourier transform of $S_3(f_1,f_2)$. It is very peaked at $\tau_1=\tau_2=0$, which means that $S_3$ is indeed frequency independent.
\end{thebibliography}
\end{document}